\documentclass{epl}
\usepackage{color}

\begin{document}

\title{Intrinsic Doping at YBCO-metal Interfaces: Quantitative Results}

\shorttitle{YBCO-metal Interfaces}

\author{U.\ Schwingenschl\"ogl
        \thanks{E-mail:\ \email{Udo.Schwingenschloegl@physik.uni-augsburg.de}}
        \and C.\ Schuster}

\institute{Institut f\"ur Physik, Universit\"at Augsburg, 
           86135 Augsburg, Germany}


\maketitle

\begin{abstract}
Charge redistribution in high-T$_c$ superconductors due to structural
defects or interfaces is known to be crucial for electronic applications
as the band structure is modified on a local scale. In order to investigate
these effects in more detail, we address the normal-state properties of
YBa$_2$Cu$_3$O$_7$ (YBCO) in the vicinity of YBCO-metal interfaces by
electronic structure calculations for well relaxed interface configurations.
Our findings can be interpreted in terms of a band-bending mechanism
complemented by local screening effects. We derive quantitative results for
the intrinsic doping of the superconducting CuO$_2$ planes due to the
metal interface. In particular, the net charge transfer amounts to 0.13
electrons in favour of each intraplane copper site, which appears to
be a typical value for interfaces of high-T$_c$ superconductors, thus
opening great possibilities for a systematic optimization of wires and
tapes from high-T$_c$ materials.
\end{abstract}

\section{Introduction}

High-T$_c$ superconductors differ strikingly from conventional
superconductors with respect to transport properties at interfaces and
grain boundaries. For example, low values of critical current densities
pose severe problems for large-current applications \cite{schneider04}.
For YBa$_2$Cu$_3$O$_{7-\delta}$, enhancement of the transport is reached
by local calcium overdoping \cite{schmehl99,hammerl00}. This effect is
connected to phenomena such as interface charging and modifications of the
local electronic band structure, which can be modelled via the charge
redistribution at superconductor-metal interfaces. In general, a large
variety of technical applications of such interfaces makes insight into
the details of the local electronic structure highly desirable.

Bending of the band structure as induced by local variations of the charge distribution
in high-T$_c$ materials is strong enough to control the superconducting
properties \cite{mannhart98,hilgenkamp02} due to large dielectric constants and small
carrier densities, characteristic for high-T$_c$ materials \cite{samara90}. As a
consequence, the Thomas-Fermi screening length, over which band-bending is effective,
reaches up to more than 1\,nm and therefore is comparable to the superconducting
coherence length. For YBa$_2$Cu$_3$O$_{7-\delta}$, experimental evidence that grain
boundaries are depleted of charge carriers comes from electron energy loss spectroscopy
\cite{browning93,babcock94}.

From the theoretical point of view, formation energies required for substituting
dopants on different YBCO lattice sites and oxygen vacancy formation energies have
been addressed by Klie {\it et al.} \cite{klie05}. Their first-principles calculations indicate
oxygen deficiency and consequently hole depletion at undoped grain boundaries, which
both are removed by calcium doping. Effects of charge modulation at the surface of
high-T$_c$ superconductors have been studied by Emig {\it et al.} \cite{emig97}, who
found that surfaces are covered by dipole layers, due to a local suppression of the gap
function. In addition, the charge imbalance arising at the boundary between a short
coherence length superconductor and a normal metal has been studied by Nikolic
{\it et al.} \cite{nikolic02} in the framework of a self-consistent microscopic approach.

The technical optimization of interfaces of high-T$_c$ superconductors calls for knowledge
about the local band-bending magnitude at the different atomic sites, which we address
in the following. Since the local electronic structure in the vicinity
of interfaces strongly depends on the local atomic configuration, it is necessary to start
from the details of the crystal structure in order to obtain quantitative results. Probably
due to their high demand on CPU time, state-of-the-art electronic structure calculations
meeting this requirement are still missing. In this letter we present
first-principles calculations for YBCO-metal interfaces, based on density functional theory
and the generalized gradient approximation as implemented in the WIEN$2k$ program
package, a full-potential linearized augmented-plane-wave code \cite{wien2k}.
Apparently for the first time, we derive quantitative results for the intrinsic doping
of the superconducting CuO$_2$ planes induced by the metal interface.

\section{Structural Considerations}

Since band-bending is proposed to take place on the length scale of the YBCO lattice
constant, the electronic properties of YBCO-metal interfaces become accessible to a supercell
approach with periodic boundary conditions. Our subsequent calculations are based on supercells
along the crystallographical $c$-axis, whereas possible $ab$-superstructures are not taken into
account for reasons of calculational complexity. The latter is justified for moderate lattice
mismatch. To be more specific, we use the YBCO $ab$-lattice constants to set up our
supercells, thus $a=3.865$\,\AA\ and $b=3.879$\,\AA\ \cite{siegrist87}. In order to
minimize forces, the atomic coordinates of the experimental YBCO structure are optimized
in a first step \cite{kouba97,kouba99}. The structural relaxation of the supercells then starts
from these data, where convergence is assumed when the boundary forces have decayed.
\begin{figure}
\oneimage[width=100mm]{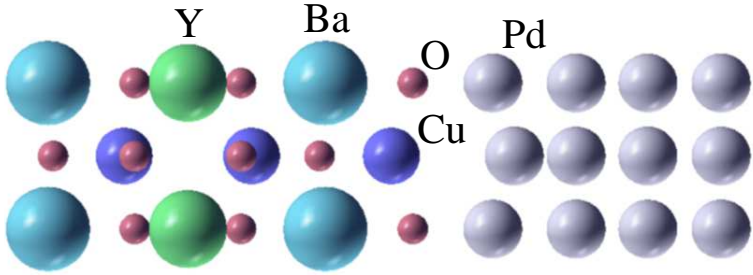}
\caption{YBCO-metal interface as resulting from the structure optimization when the Pd sites
are assumed to be placed on top of the interface O sites (configuration A)}
\label{fig1}
\end{figure}

As metallic substituent we choose fcc palladium due to a minimal lattice mismatch
($c_{\rm fcc}=3.89$\,\AA) of about 0.7\%. Along the $c$-axis, our supercells consist of  
2 YBCO unit cells, terminated by the CuO-chain layers 
\cite{xin89,edwards92,derro02}, and 4 metallic unit cells in an [001] orientation.
The resulting supercell therefore belongs to space group Pmm2, comprising 46 inequivalent
atomic sites. As starting point of the structural relaxation, there are two natural interface
configurations where the metallic sites are placed on top of the interface oxygen (configuration
A) or copper (configuration B) sites, respectively. Since we found that our conclusions
about the band-bending mechanism do not depend on the configuration used, the
following considerations focus on configuration A. As far as these results are
concerned, $ab$-superstructures thus likewise seem to play a minor role.

The structure optimization of our supercells shows a strong tendency towards
Pd-O bonding at the interface. In addition, significant repulsion of copper and palladium
is observed.  In the metal region, the structural relaxation affects almost exclusively the
first atomic layer, whereas in the YBCO region mainly the CuO-chain layer as well as
the BaO layer are involved. The O sites of the interface CuO-chain are shifted out of plane.
In configuration A, a strong buckling of the first Pd layer allows the system to avoid
Cu-Pd overlap, whereas this buckling is considerably reduced in configuration B. The
supercell of configuration A as resulting from the structure optimization is depicted in Fig.\ \ref{fig1}
in a projection along the $a$-axis, thus perpendicular to the CuO-chains. Corresponding
bond lengths are summarized in Tab.\ \ref{tab1}. While the bond lengths
in the Cu-O chain of bulk YBCO amount to 1.94\,\AA, the values 2.02\,\AA\ and
2.08\,\AA\ are found for interfaces A and B, respectively. Moreover, we observe a
reduced bond length of 1.81\,\AA\ between the chain Cu sites and the
neighbouring O sites in the BaO layer of configuration B, whereas in configuration A
the bulk value of 1.90\,\AA\ persists.
\begin{table}
\begin{tabular}{l|c|c}
&Configuration A & Configuration B \\\hline
d$_{\rm Cu-Pd}$&3.32\,\AA, 3.79\,\AA&2.68\,\AA, 3.81\,\AA\\
d$_{\rm O-Pd}$&2.12\,\AA, 3.87\,\AA&2.70\,\AA, 2.72\,\AA\\
d$_{\rm Cu-O}$&2.02\,\AA&2.08\,\AA\\
d$_{\rm Cu-O_{Ba}}$&1.90\,\AA&1.81\,\AA
\end{tabular}
\vspace{0.5cm}
\caption{Bond lengths close to the YBCO-metal interface as resulting from the structure
optimization.}
\label{tab1}
\end{table}

\section{Results}

In the following, band structure data for bulk YBCO, without interface, will act as
a reference for the interpretation of the data. We mention
that our bulk YCBO band structure agrees well with previous calculations using the
local density approximation and likewise reveals good agreement with experimental
observations, see \cite{pickett89,pickett90,wechsler97} and the references given therein.

\begin{figure}
\oneimage[width=65mm]{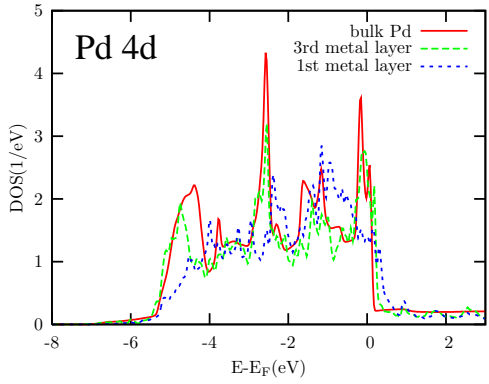}
\caption{Partial Pd $4d$ DOS for metal layers parallel to the YBCO-metal interface.
With increasing distance to the interface the curves quickly converge to the bulk Pd DOS.}
\label{fig2}
\end{figure}
For the YBCO-metal interface, we first investigate the palladium sites.
Fig.\ \ref{fig2} displays partial Pd $4d$ densities of states for the sites
in the 1st and 3rd metal layer. Here and in the following, layers are counted with respect
to the interface, i.e.\ the 1st metal layer is attached to the CuO-chains of the YBCO cell.
For comparison, we include a fcc bulk Pd DOS in Fig.\ \ref{fig2}. While for the 1st and
2nd metal layer Pd states are shifted
from the Fermi level both to higher and lower energies, the local DOS for sites in the
3rd and further layers almost resembles the bulk DOS. Hence the
electronic screening length is found to be small in the palladium region, as expected.

\begin{figure}
\twoimages[width=65mm]{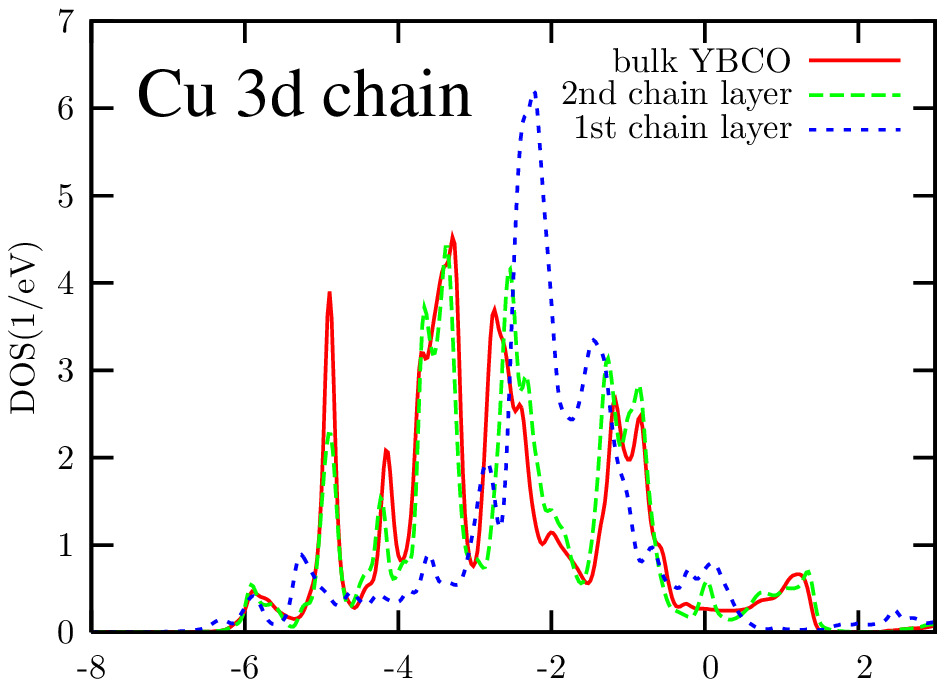}{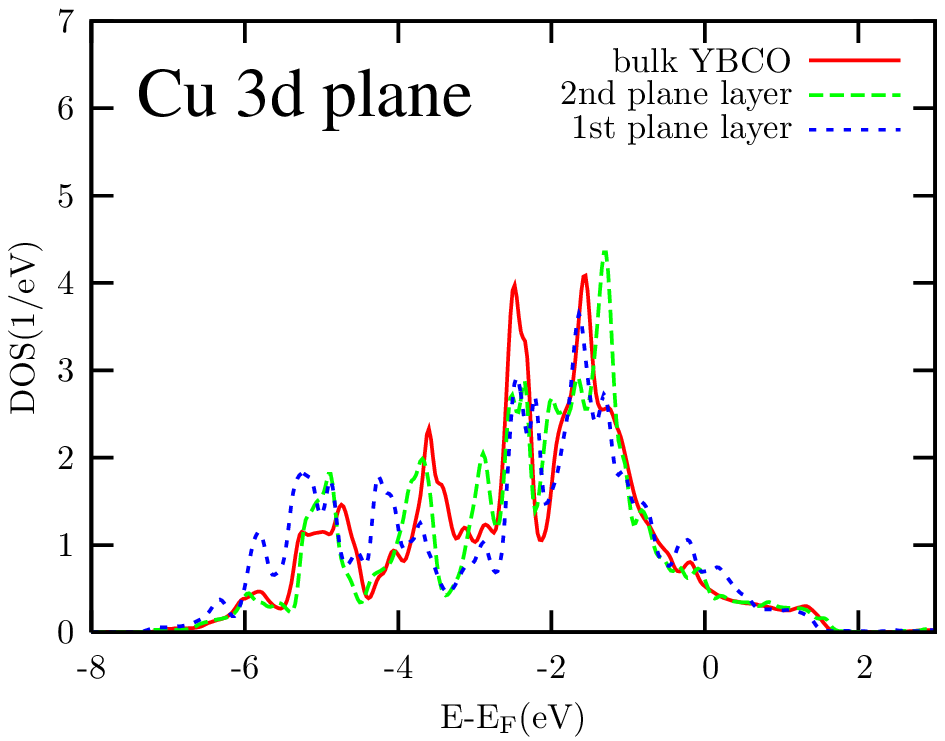}
\twoimages[width=65mm]{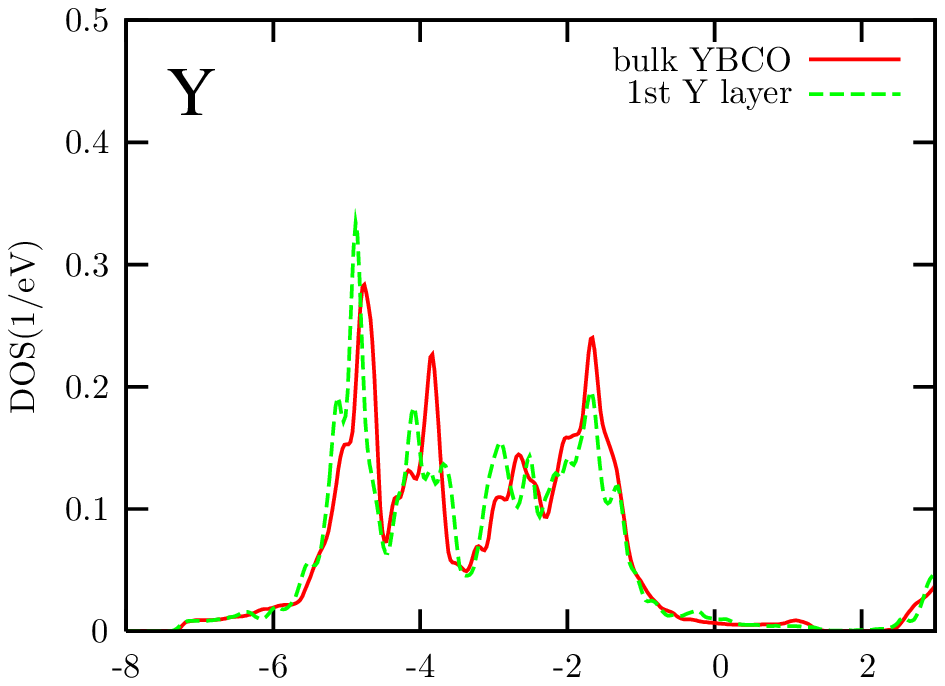}{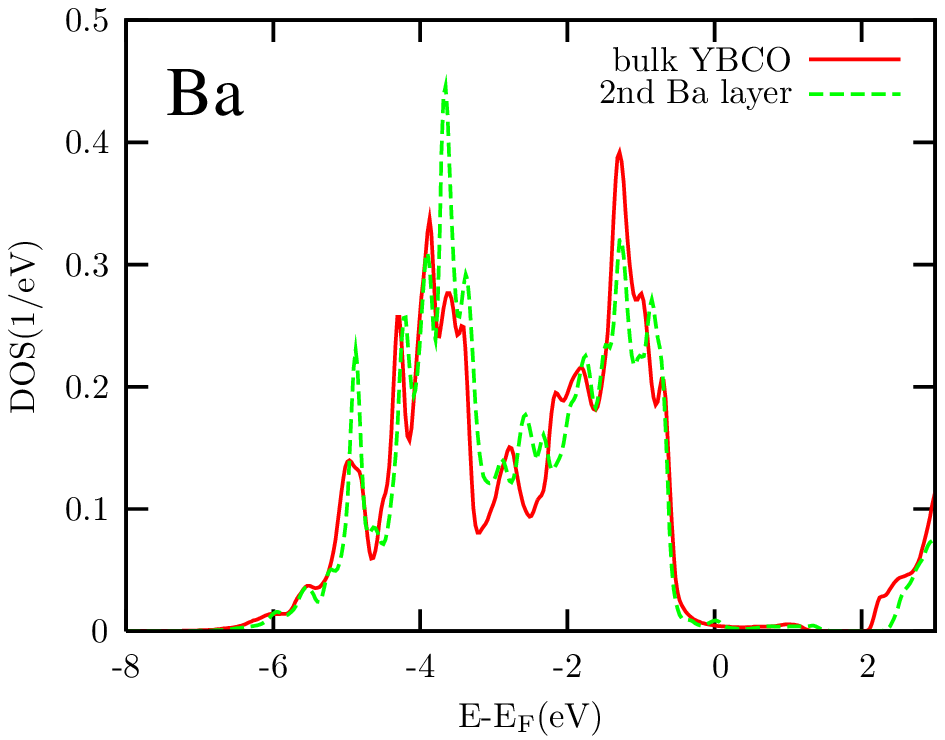}
\caption{Partial Cu $3d$ (CuO-chain and CuO$_2$-plane sites), Y and Ba DOS. The second CuO-chain layer
resembles the bulk YBCO Cu DOS when the latter is shifted by 0.3\,eV to lower energies.
Similar energetical shifts of 0.3\,eV and 0.45\,eV are found for the Y and Ba sites, respectively.
The second CuO$_2$-plane
resembles the bulk YBCO Cu DOS in the vicinity of the Fermi level when the latter is shifted
by 0.2\,eV to lower energies, which corresponds to a reduction of 0.13 holes per Cu site.}
\label{fig3}
\end{figure}
Turning to the YBCO region, we first address the CuO-chain layers, for which Cu $3d$
densities of states are shown in the upper panel of Fig.\ \ref{fig3}. Because of Cu-O
bonding, copper and oxygen states give rise to a broad structure in the energy region from
about $-6.5$\,eV to 1.5\,eV, with respect to the Fermi level. While the 1st chain layer
is affected by strong modifications of the electronic bands due to structural relaxations
at the interface, the Cu sites in the 2nd chain layer (one YBCO unit cell off the interface)
are found to resemble the electronic structure of the respective bulk YBCO Cu sites,
see Fig.\ \ref{fig3}. However, we note that the bulk DOS is shifted to lower energies by
0.3\,eV in order to reconcile the curves. As the fine structure of the 2nd chain layer DOS
is hardly affected by the interface, we can interpret this shift in terms
of almost ideal down bending of the electronic bands due to a modified Fermi level.

Small contributions of Y and Ba states in the copper dominated energy interval, as shown
in the middle and lower panel of Fig.\ \ref{fig3}, respectively, are a consequence of
weak covalent bonding, mainly with oxygen states. Although Y and Ba act as
electron donors, in principle, these contributions close to the Fermi level are well suitable for analyzing
possible band-bending effects at the cation sites. For the 1st Y and 2nd
Ba layer (the 1st Ba layer is affected by structural recombinations) the DOS
curves of Fig.\ \ref{fig3} show almost perfect agreement with the respective
bulk YBCO Y/Ba DOS when adequate energetical shifts are applied. Again, for the Y sites
a shift of 0.3\,eV is convenient, while a shift of 0.45\,eV is necessary
for the Ba sites. To understand this fact, effects of
electronic screening have to be taken into consideration. Both the Y and Cu chain sites (see the
previous discussion) reveal a finite number of electronic states right at the Fermi level,
whereas for Ba the DOS almost vanishes. For this reason, screening is less efficient at
the Ba sites and band-bending effects become more pronounced. Energetical shifts
of core levels at atomic sites near the interface are fully consistent with this interpretation. 

The local charge distribution in the CuO$_2$-planes near interfaces is of central interest for
the YBCO transport properties. In particular, effects of interface charging and modifications of the
electronic band structure near grain boundaries, where superconductivity is locally suppressed due
to the imperfections of the crystal structure, can be modelled. Partial Cu $3d$ densities of
states for CuO$_2$-plane sites in the 1st and 2nd layer off the interface are compared to the
corresponding bulk YBCO Cu DOS in Fig.\ \ref{fig3}. While for the 1st layer some additional Cu
states appear around the Fermi energy, which grow out of the modified Cu-O bonding at
the interface, the DOS of the 2nd CuO$_2$-plane perfectly fits the bulk YBCO Cu DOS shape
close to the Fermi level. In contrast to the local band-bending magnitudes at the
previously addressed YBCO sites, here an energetical shift of 0.2\,eV is found to be sufficient
to reconcile the curves. This fact agrees well with the screened band-bending mechanism
due to the high charge carrier density in the CuO$_2$-planes. As compared to the bulk
charge distribution, the 0.2\,eV down bending of the Cu $3d$ bands comes along with
a charge carrier depletion of 0.13 holes per Cu site. Importantly, both this value and
the calculated band-bending magnitudes seem to be rather general results as similar numbers
are found when silver is used as metallic substituent.

Because of an electrostatic screening length of a few nanometers in
high-T$_c$ cuprates and the inhomogeneity of the crystal structure, the
conventional band-bending models based on a continuum description of the charge
distribution are not applicable. Screening effects result in a significant reduction of the
band-bending magnitude and therefore in reduced charge transfer. In particular, the
calculated 0.13 hole depletion per Cu site seems to explain the observation that
local hole-doping of grain boundaries, via replacing Y by Ca, leads to enhanced supercurrent
densities. Schmehl {\it et al.} \cite{schmehl99}
find the maximal enhancement for 30\% Ca doping, i.e. doping of 0.3 holes per CuO$_2$
unit. Assuming that doping effects predominantly affect the CuO$_2$-planes adjacent
to the Y/Ca sites, this value agrees even quantitatively with our band structure result of
0.26 holes. The latter maintains when we take into account that the
experimental value relies on oxygen deficient samples, YBa$_2$Cu$_3$O$_{7-\delta}$,
since the magnitude of screened band-bending in the CuO$_2$-planes is expected to be
almost independent of the oxygen content.

Moreover, our findings cast doubts on the role of oxygen deficiency reduction by Ca doping
as claimed by Klie {\it et al.} \cite{klie05}. The authors attribute the reduction of the
local hole depletion at grain boundaries to the removement of intrinsic excess oxygen vacancies,
induced by Ca doping. In contrast, our results indicate that modifications of the oxygen concentration
are less important. At least for YBCO-metal interfaces band-bending is sufficient to
explain the reported doping effects.

Xu and Ekin report on specific resistivities for YBCO-Au interfaces of
$10^{-4}\,\Omega\,{\rm cm}^2$ to $10^{-3}\,\Omega\,{\rm cm}^2$ at low
temperatures \cite{xu04}. However, our calculations for normal-state YBCO interfaces
do not show a significant reduction of the Cu $3d$ DOS in the vicinity of the Fermi level.
Even though no insulating layer is formed at the interface, the screened band-bending
mechanism still can explain the observed interface resistivity. Since the charge carriers in
the CuO$_2$-planes are reduced, the superconductivity is locally disturbed.

\section{Conclusions}

We have discussed electronic structure calculations for interfaces between the short
coherence length superconductor YBa$_2$Cu$_3$O$_7$ and the normal metal palladium
in order to analyze the influence of the charge redistribution induced by interfaces
on the local doping of the superconductor. Based on well relaxed interfaces, we have
found that modifications of the charge distribution with respect to the bulk are well
screened in the metal region, whereas they affect the superconductor on a nanometer length scale.
Specifically, our calculations result in a net charge transfer of about 0.13 electrons
in favour of each Cu site in the CuO$_2$-planes, in correspondence with the experimental
observation of charge carrier depletion. 

In particular, our results are consistent with the fact that artificial
hole-overdoping of grain boundaries yields significant improvement of the
transport properties. The depletion of charge carriers due to the interface
is compensated and a state of optimal doping is reached. Our calculations agree
even quantitatively with experiment as concerns the amount of hole-overdoping
needed for obtaining an optimally doped interface. It turns out that
screening strongly affects the magnitude of the charge redistribution
in the CuO$_2$-planes, which has serious consequences for the design of
interface structures, such as S-N-S Josephson contacts, for instance.
Expectedly, the mechanism of screened band-bending and the net charge
transfer are almost independent of the specific high-T$_c$ material and
metal forming the interface. Therefore our results are very general and
can be applied to a large variety of interface configurations, thus
paving the way for systematic material optimization.

\begin{acknowledgments}
This work was supported by the Deutsche Forschungsgemeinschaft through SFB 484.
We gratefully acknowledge valuable discussions with U.\ Eckern, V.\ Eyert, J.\ Mannhart,
and T.\ Kopp.
\end{acknowledgments}

\end{document}